\documentclass[mathleft %,draft
]{an}
\usepackage{graphicx}
\usepackage{lipsum}
\usepackage{times}
\usepackage{url}
\usepackage{natbib}
\bibpunct{(}{)}{;}{a}{}{,}
\begin{document}

% The following seven commands are intended for editorial usage and should be ignored by
% the author(s).
\Pagespan{789}{}% Document's page range. 
% If second parameter is left empty, the last page is computed automatically.
\Yearpublication{2006}%
\Yearsubmission{2005}%
\Month{11}%   
\Volume{999}%  
\Issue{88}% 
% \DOI{This.is/not.aDOI}% 

\title{A quest for activity cycles in low mass stars}

\author{K. Vida\thanks{Corresponding author:
  \email{vidakris@konkoly.hu}\newline},
L. Kriskovics, 
K. Ol\'ah
%\inst{1}
%Example 
%for footnote, note the usage of the \texttt{fnmsep}
%command as separator between institute number and footnote mark} 
}
\titlerunning{A quest for activity cycles in low mass stars}
\authorrunning{K. Vida}
\institute{
Konkoly Observatory of the Hungarian Academy of Sciences
H-1121 Budapest, Konkoly Thege Mikl\'os str. 15-17.
}
\received{30 May 2005}
\accepted{11 Nov 2005}
\publonline{later}


\abstract{
Long-term photometric measurements in a sample of ultrashort-period ($P\approx0.5$ days or less) single and binary stars of different interior structures are analysed. A loose correlation exists between the rotational rate and cycle lengths of active stars, regardless of their evolutionary state and the corresponding physical parameters. The shortest cycles are expected on the fastest rotators in the order of 1-2 years, which is reported in this paper.
}

\maketitle

\section{Introduction}

The 11 year-long activity cycle on the Sun is known for a long time. On other stars \cite{1978ApJ...224..182P} were the first to find long-term brightness changes which were interpreted as starspot cycles. Using long-term photometric measurements, \cite{2000AAA...356..643O} found that other stars can also have multiple activity cycles -- similar to the Gleissberg and Suess cycles on the Sun -- and that the length of the cycles vary in time. 

\cite{1978ApJ...226..379W} and \cite{1995ApJ...438..269B} 
studied the chromospheric activity of F--M-type stars using long-term Ca\,H\&K data from the Mt. Wilson survey, and detected cyclic variations in the H\&K flux. Different relations are known between the rotation period, the length of the activity cycle, and other stellar properties.
Using the Mt. Wilson survey data, \cite{1996ApJ...460..848B} found that there is a connection between the cycle length and the rotation period, namely, that faster rotating stars have shorter activity cycles. 
The authors also gave an explanation for the cycle lengths of different stars using dynamo theory.
 \cite{1993ApJ...414L..33S} proposed $(P_\mathrm{cyc}/P_\mathrm{rot})^2$, the square of the ratio of the cycle length and the rotation period as a quantity to parametrize activity cycles. \cite{Brandenburg:1998dt} and \cite{1999ApJ...524..295S} found a dependence between the cycle and rotational frequencies $\omega_\mathrm{cyc}/\Omega$ and the Rossby-number.
\cite{2002AN....323..361O} studied photometric data of active stars, and found a similar correlation between the rotation and cycle length, as \cite{1996ApJ...460..848B}.

Therefore, if we seek activity cycles, it is effective to monitor fast-rotating objects, i.e., cycles may be found after a few years of monitoring already, contrary to the case of stars with longer rotation period.
We present the analysis of four objects, all of them having a rotation/orbital period in the order of 0.5 days: EY Dra, V405\,And,  GSC 3377-0296, and V374 Peg.

EY Dra is a single, active dM1-2 star that was thoroughly analyzed lately by \cite{2010AN....331..250V}. The star showed slow evolution on monthly time scale, and possibly a flip-flop mechanism. The activity cycle presented here was reported also in that paper.

V405\,And is a grazing eclipsing binary which was studied in detail first by \cite{1997AAA...326..228C}. They found an orbital period of $P=0.465$ days and a small eclipse in the light curve. The  spectral type of the components were also determined: M0V and M5V were found for the primary and the secondary, respectively. Both components were showed strong emission in the region of the $H\alpha$  line.
\cite{2009AAA...504.1021V} analyzed the system using  $BV(RI)_C$  photometric measurements and spectroscopic data. Strong flares in the light curve and the $H\alpha$ region were observed, and by modeling the system they showed that the radius of the primary was significantly larger than the theoretically predicted value, while the size of the secondary fit the mass-radius relation. This result was later confirmed by \cite{2011AJ....142..106}, who used an independent spectroscopy-based method to determine the size of the components. 

GSC 3377-0296 was studied to date only by \cite{2007IBVS.5772....1L}, who identified the object as a heavily spotted RS CVn system with an orbital period of 0.422 days. 

V374 Peg is a single, fully convective M4 dwarf with $0.28M_\odot$ \citep{2000AAA...364..217D}. According to \cite{2006Sci...311..633D} and \cite{2008MNRAS.384...77M} , the star has very weak differential rotation, and a stable, poloidal, axisymmetric magnetic field. 

In this paper new, long-term datasets of these objects are presented in the hope of finding activity cycles. 

\begin{figure*}
\centering
\includegraphics[width=0.325\textwidth]{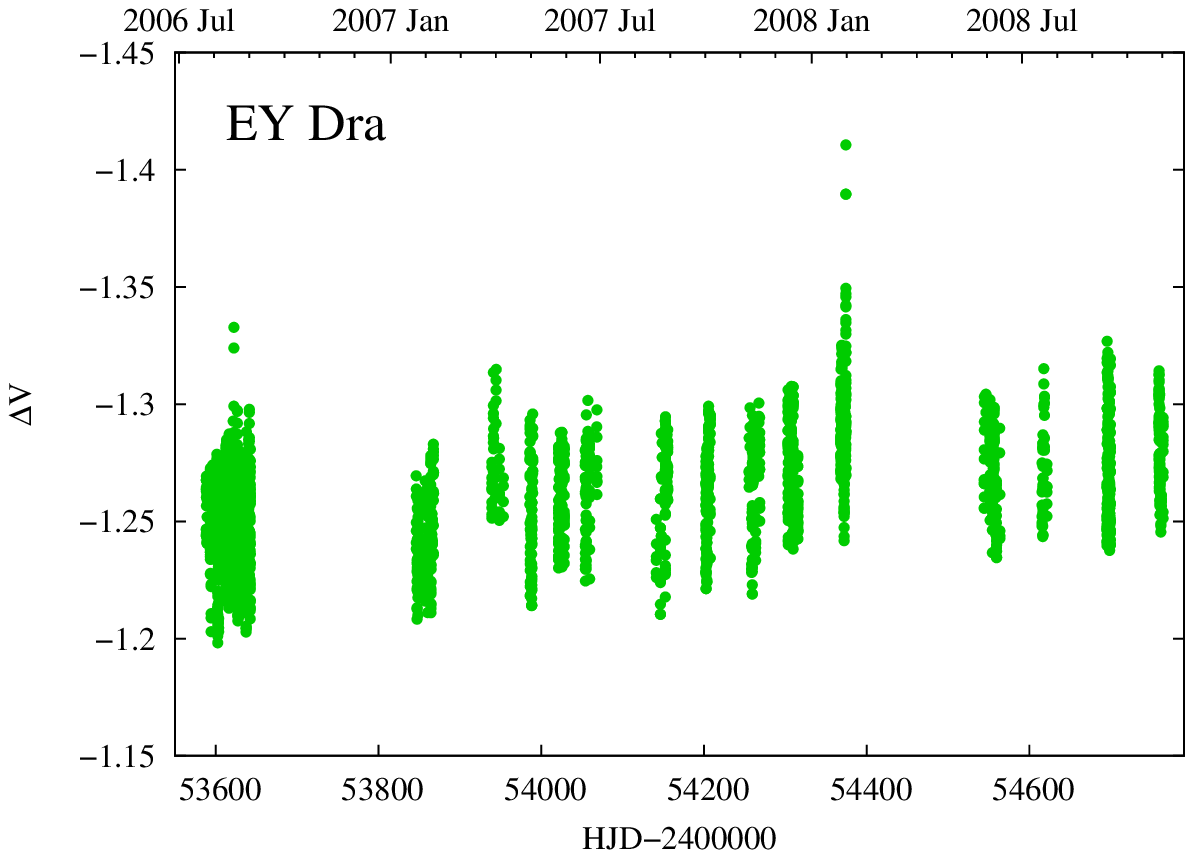}
\includegraphics[width=0.325\textwidth]{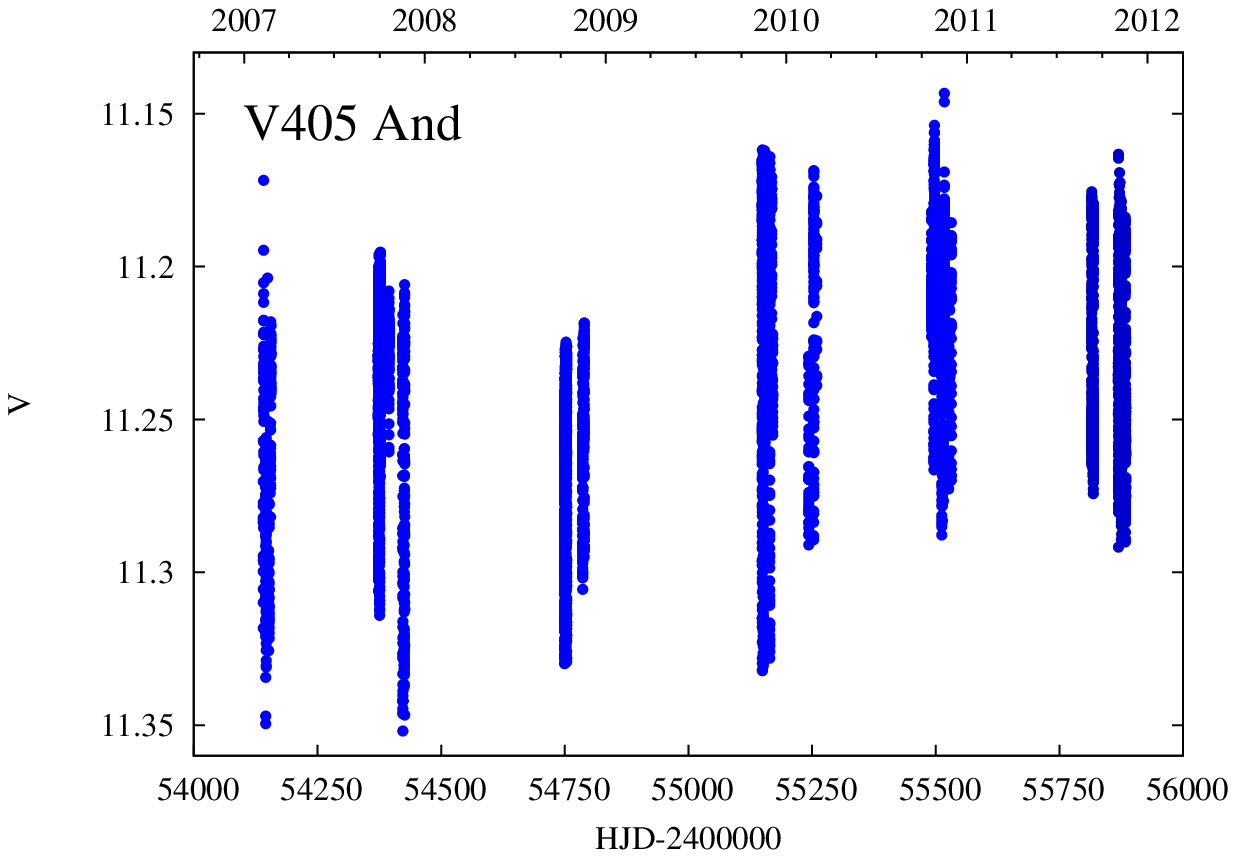}
\includegraphics[width=0.325\textwidth]{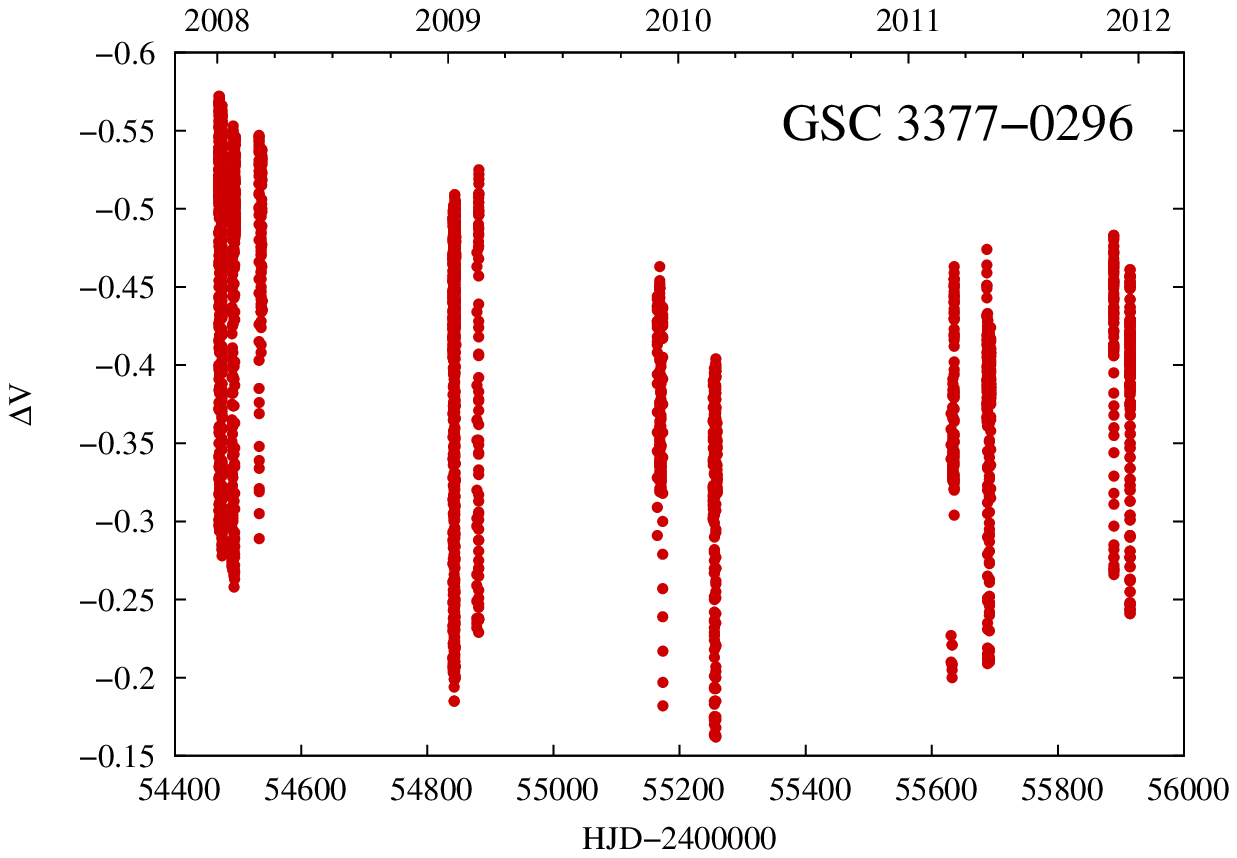}

\includegraphics[width=0.325\textwidth]{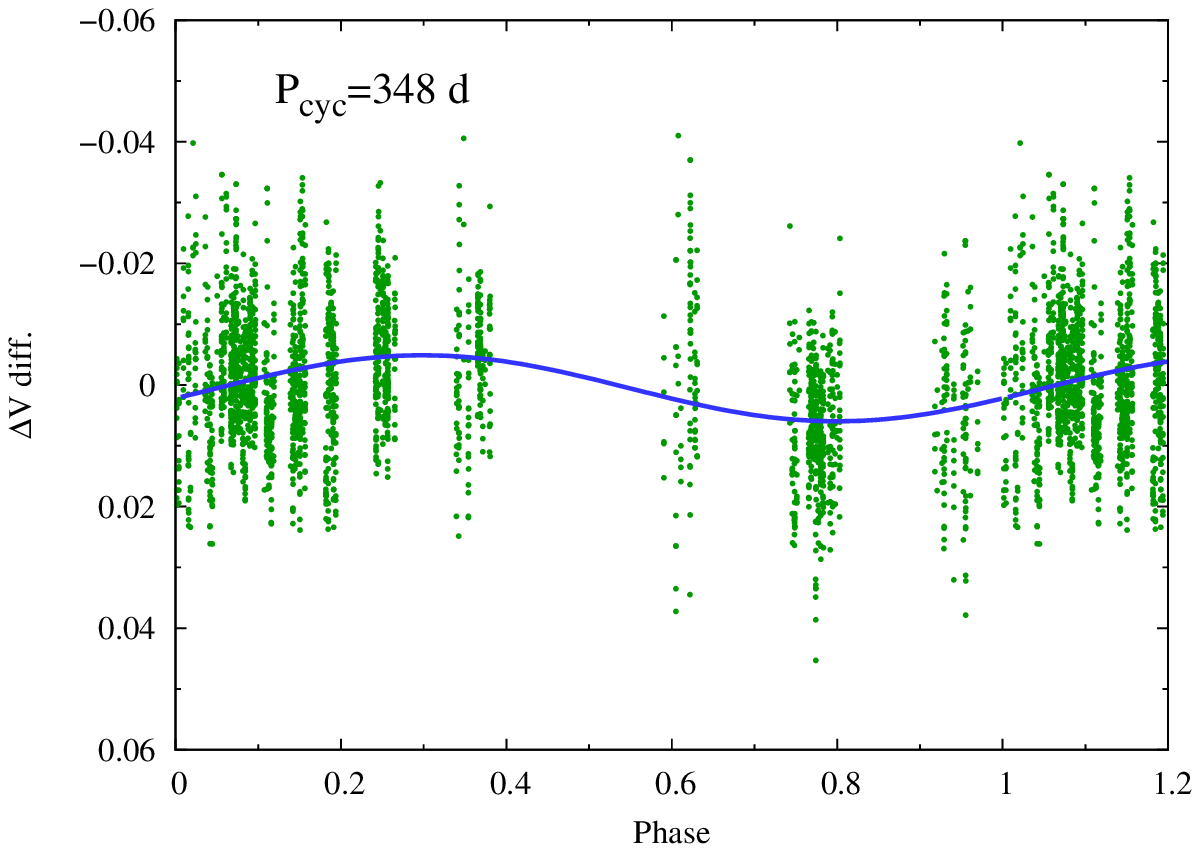}
\includegraphics[width=0.325\textwidth]{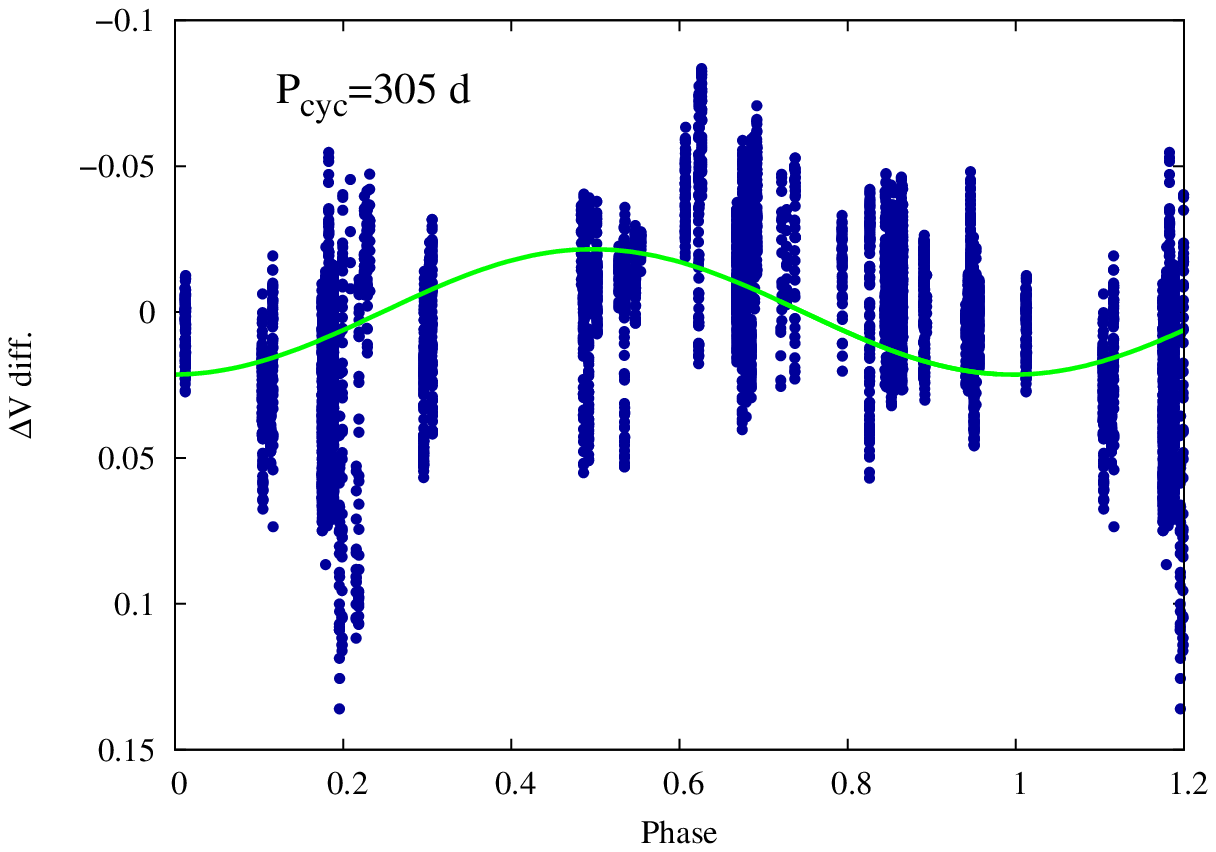}
\includegraphics[width=0.325\textwidth]{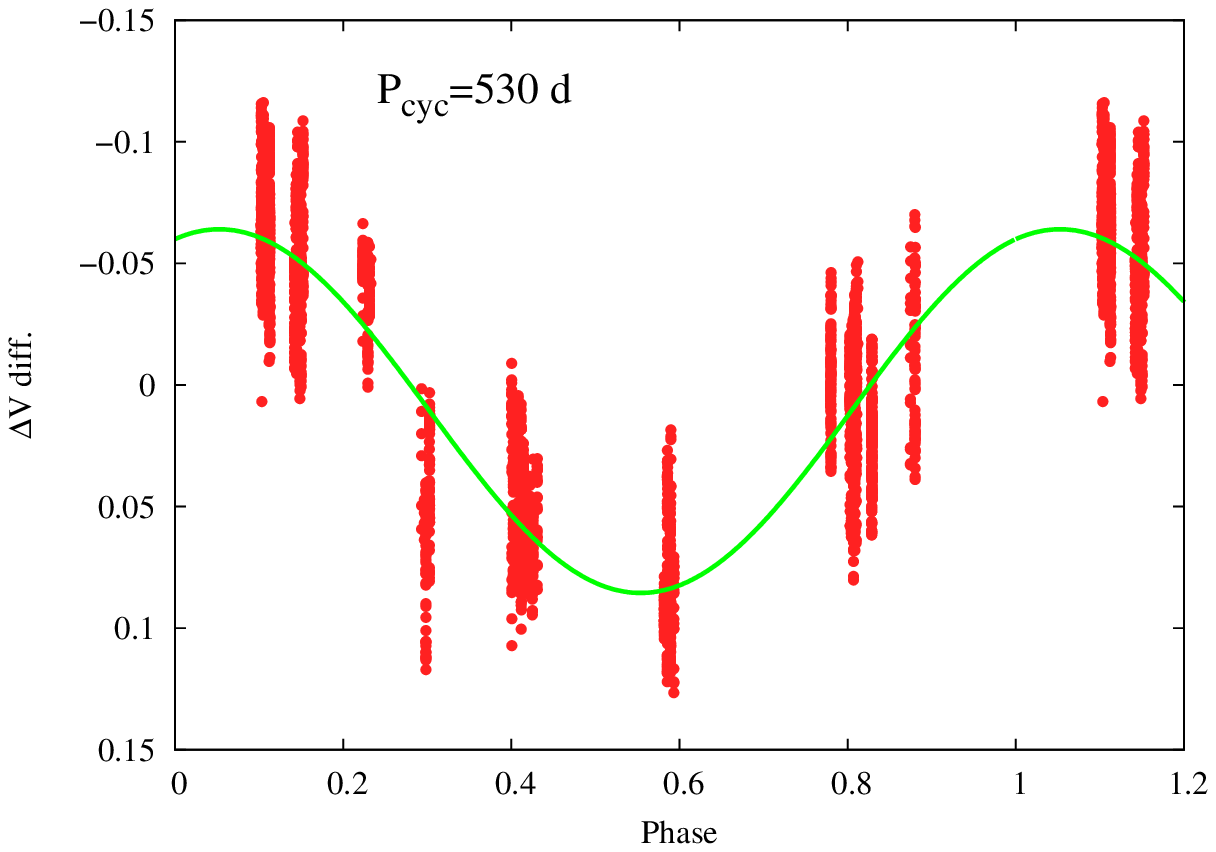}

\includegraphics[width=0.325\textwidth]{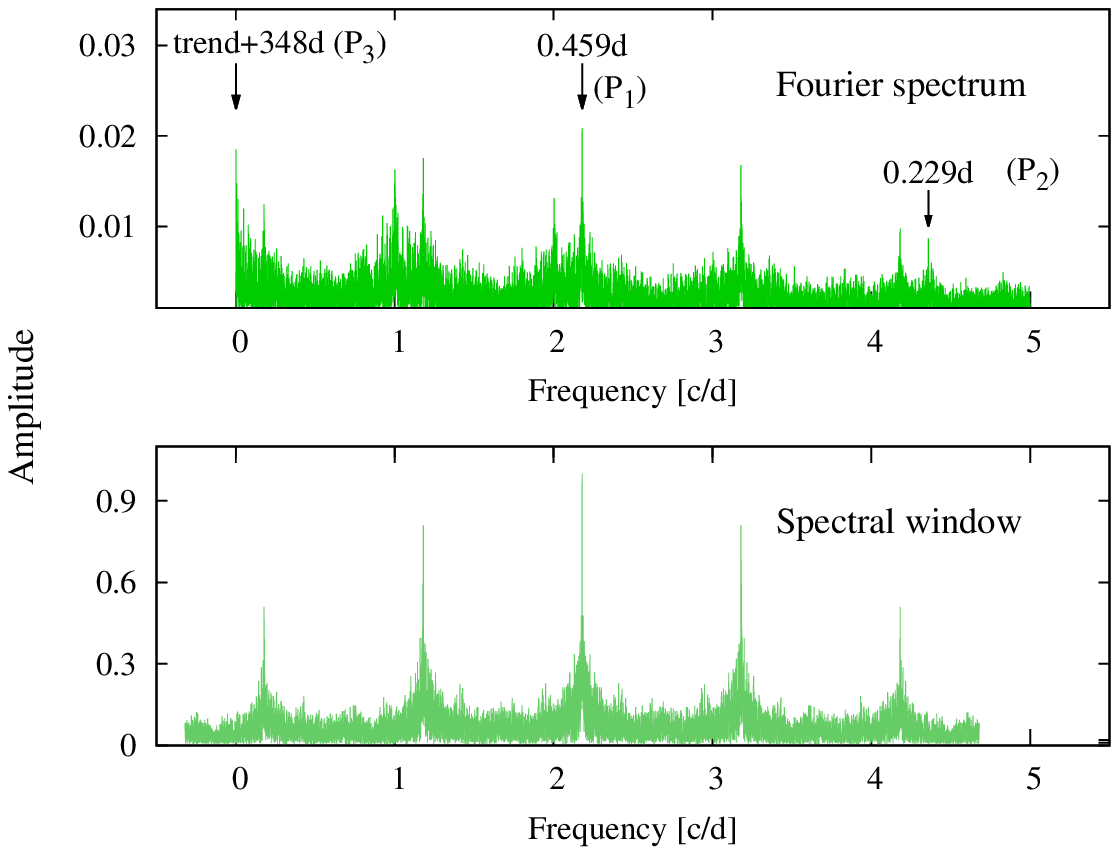}
\includegraphics[width=0.325\textwidth]{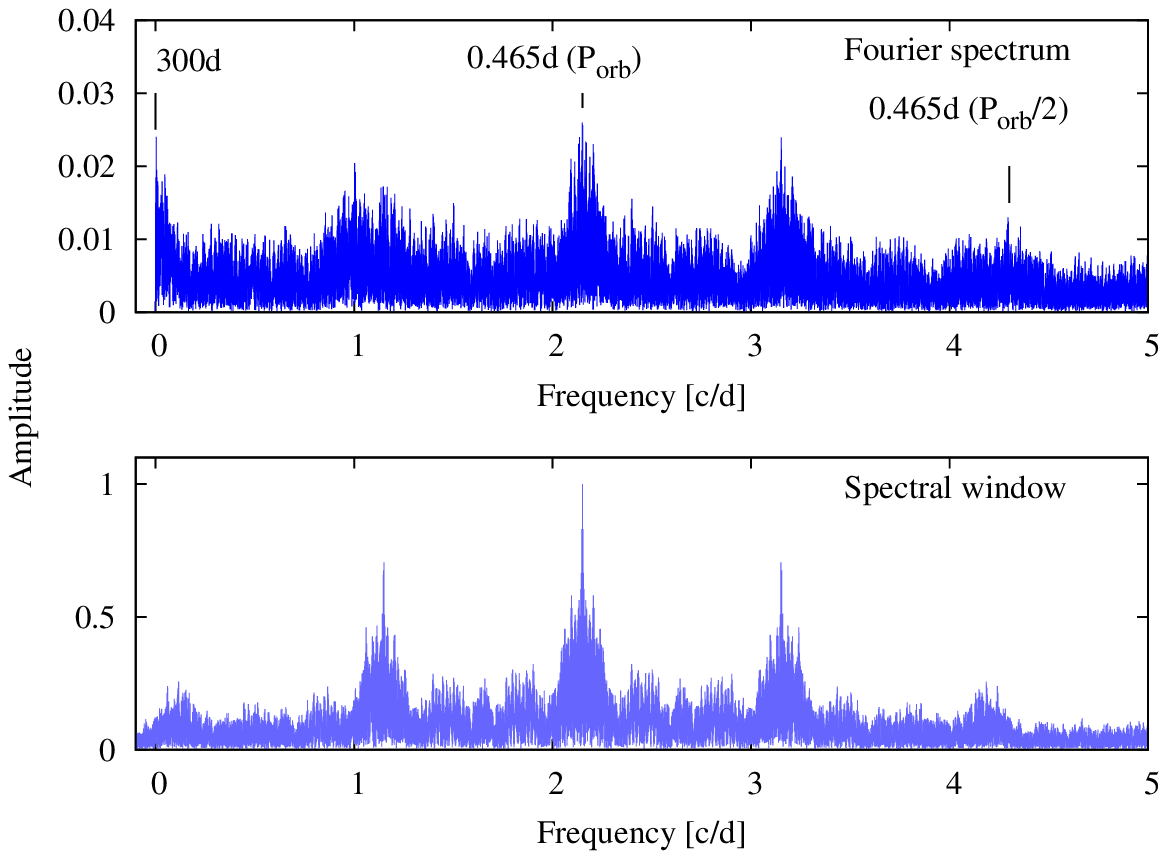}
\includegraphics[width=0.325\textwidth]{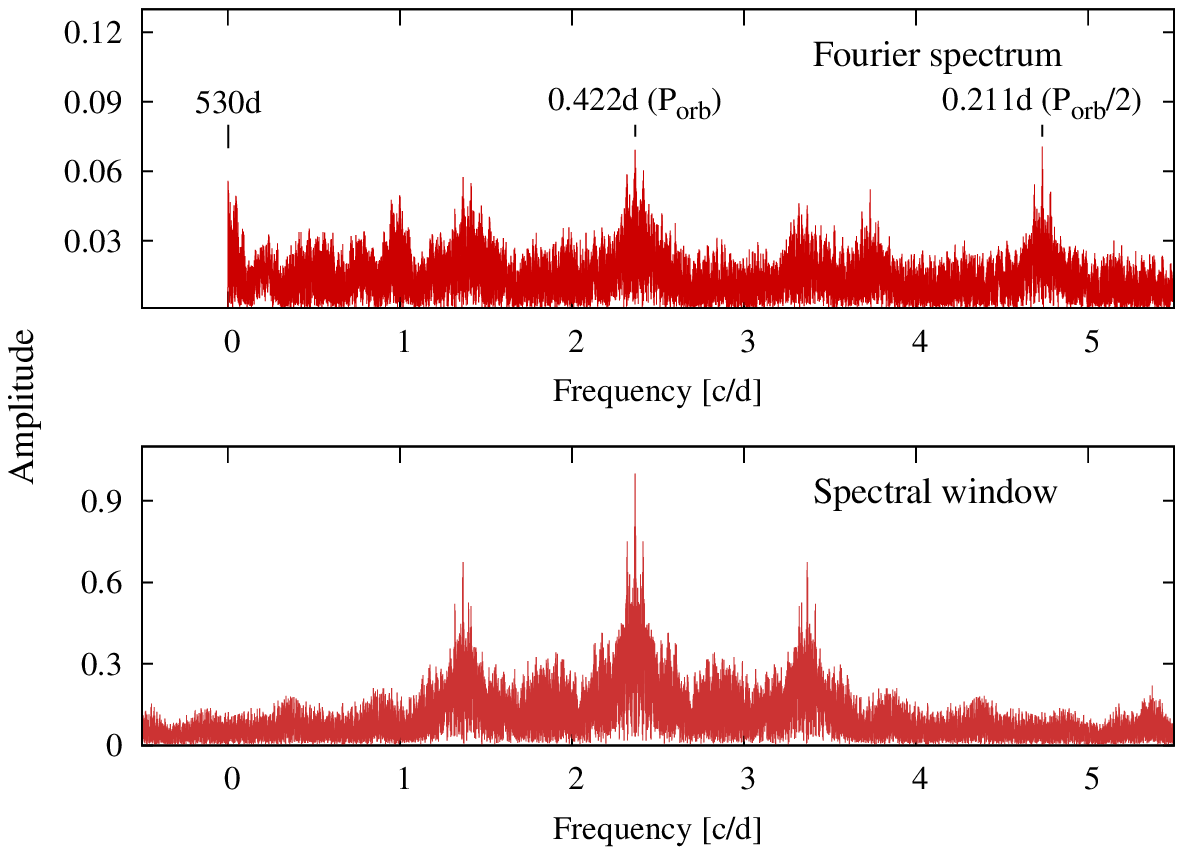}

\caption{\textit{From top to bottom:} Observed light curves, phased light curves with the activity cycle after pre-whitening with the rotational modulation, and Fourier spectra of the targets with the spectral windows. The plots from left to right show data of EY Dra, V405\,And, and GSC 3377-0296.}
\label{fig:figure}
\end{figure*} 

%%%%%%%%%%%%%%%%%%%%
\section{Observations}
Data of EY Dra were obtained using the 60\,cm telescope of the Konkoly Observatory at Sv\'abhegy, Budapest equipped with a Wright Instruments $750\times1100$ CCD camera. V405\,And was observed both by the 60\,cm telescope, and on Piszk\'estet\H{o} Mountain station of the Konkoly Observatory, using the 1m RCC telescope and a Princeton Instruments $1300\times1300$  CCD. The other targets were monitored only by the 1m RCC telescope. The observations presented in this paper were obtained using Johnson $V$ filter. The datasets cover time ranges of 3--5 years  (the exact values are shown in Table \ref{tab:table}). Data reduction was done using standard IRAF\footnote{IRAF is distributed by the National Optical Astronomy Observatory,
which is operated by the Association of Universities for Research in Astronomy, Inc., under cooperative agreement with the National Science
Foundation.} 
procedures. Differential aperture photometry was done using the DAOPHOT package.

%%%%%%%%%%%%%%%%%%%%
\section{Analysis \& discussion}

\begin{figure}
\centering
\includegraphics[width=0.48\textwidth]{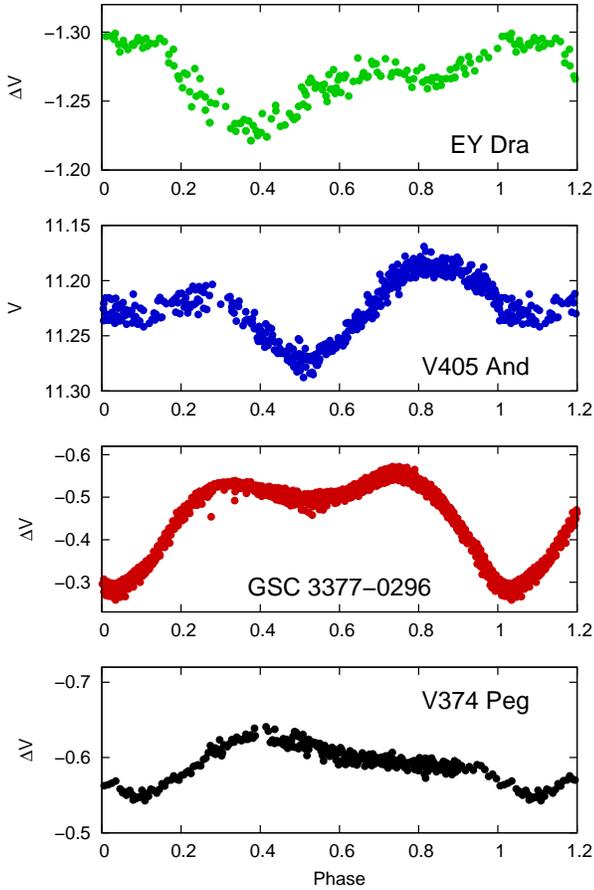}
\caption{\textit{From top to bottom:} Typical phased light curves of EY Dra, V405\,And, GSC 3377-0296, and V374 Peg.}
\label{fig:phasedlc}
\end{figure}

\begin{figure}
\centering
\includegraphics[width=0.45\textwidth]{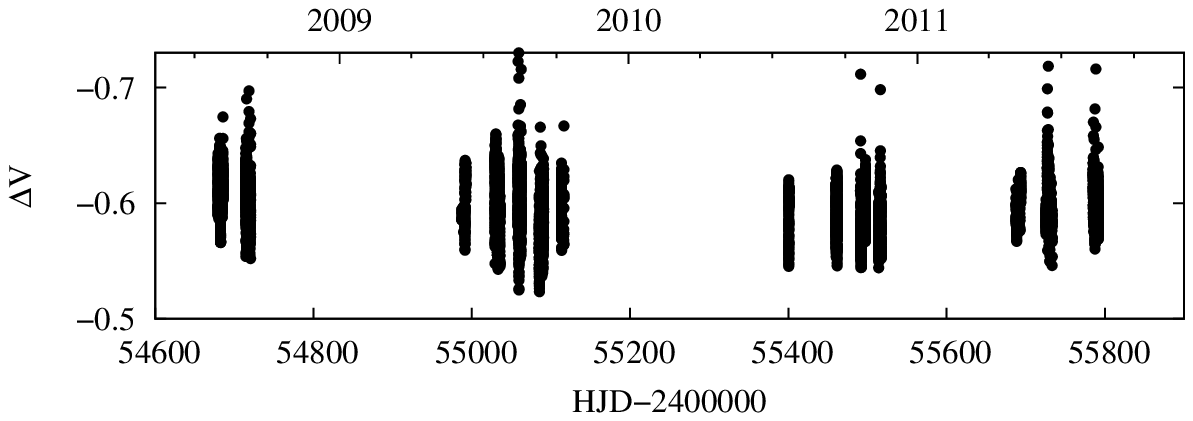}
\includegraphics[width=0.45\textwidth]{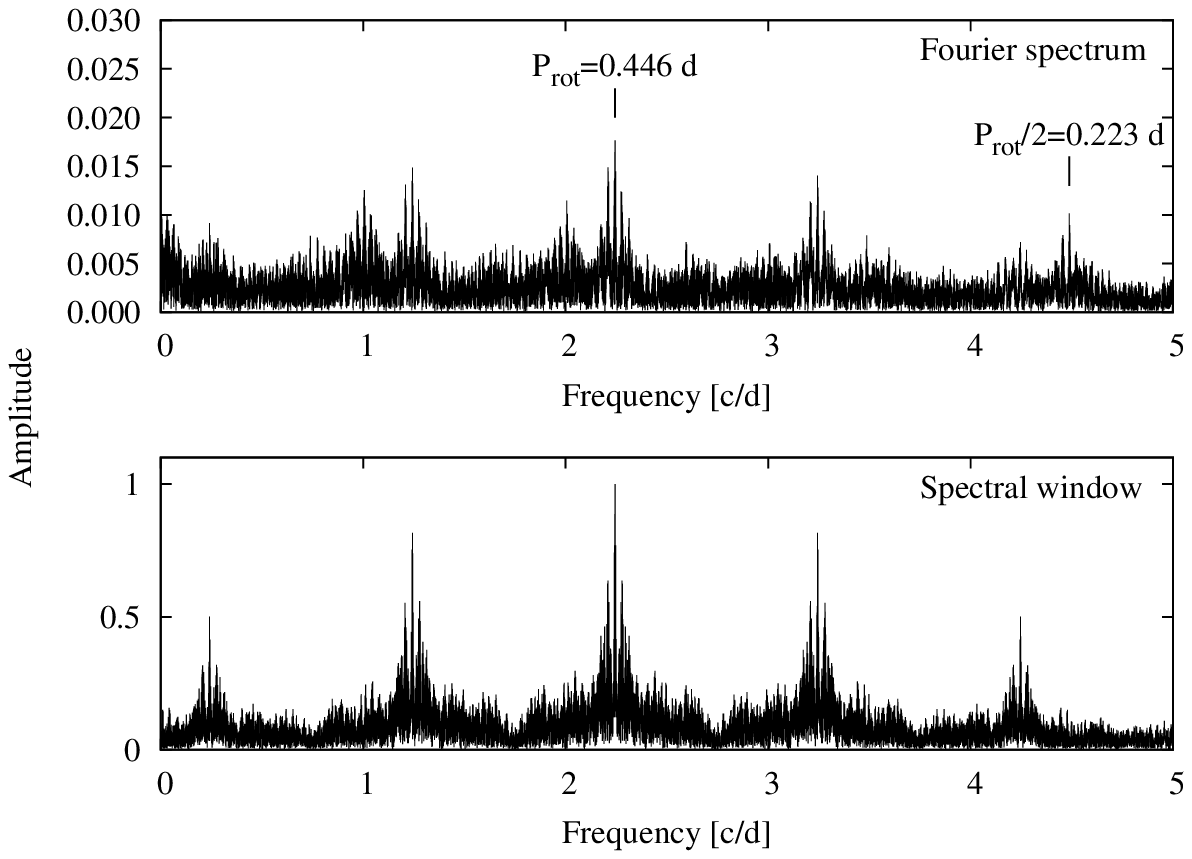}
\caption{\textit{From top to bottom:} $V$ light curve, Fourier-spectrum and spectral window of V374 Peg. Note, that the top plot is truncated, the eruptions caused by strong flares cannot be seen.}
\label{fig:v374peg}
\end{figure}

\begin{figure}
\centering
\includegraphics[width=0.48\textwidth]{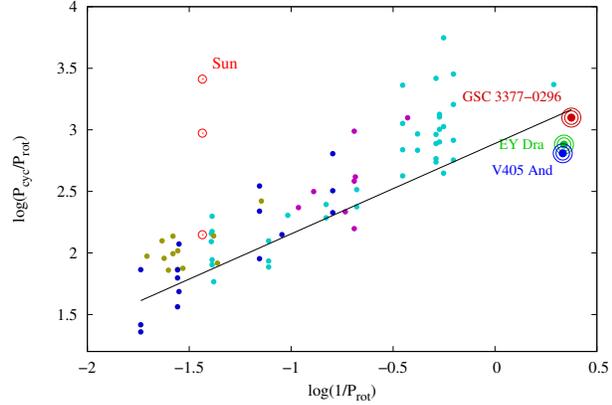}
\caption{Rotation period--activity cycle relation, based on \cite{2002AN....323..361O}. Different colors denote data from different surveys. The line shows the fit to the shortest cycles.}
\label{fig:2002AN....323..361O}
\end{figure}

\begin{table}
\caption{The basic properties of the objects, the lengths of the detected activity cycles, their significance, and the total length of the observed light curves.}

\begin{tabular}{r|ccc|c}
			&EY		&	V405		&	GSC 3377-&	V374 \\
			&Dra		&	And		&	0296		&	Peg\\
\hline
Sp. type		&dM1-2	&	dM0+dM5	&	K3?		&	dM4\\
Binarity		&--		&	+		&	+		&	--\\
Mass ($M_\odot$)&0.5	&0.49+0.21	&	?		&	0.28\\
$P_\mathrm{rot} (d)$	&0.459 	&	0.465&	0.422 &0.445 \\
$P_\mathrm{cyc} (d)$&348	&	305 	&	530 		&	--\\
Significance &5.6$\sigma$ 	& 6.1$\sigma$ 	& 8.4$\sigma$&--\\
LC length $(d)$ & 1176 	& 1391 	& 1446 		& 1112\\
\end{tabular}
\label{tab:table}
\end{table}

The Fourier-analysis of the observed light curves was done using the MUFRAN software \citep{1990KOTN....1....1K}. The Fourier-spectra of the data with the spectral windows of EY Dra, V405\,And, and GSC 3377-0926 are plotted in Fig. \ref{fig:figure}. Before determining the length of the starspot cycle the light curves were first pre-whitened with the rotational modulation. The main feature on all Fourier-spectra is the rotational modulation and a signal at the double frequency (and their aliases), as all targets have typically two dominant active regions on their surface (see Fig. \ref{fig:phasedlc} with the phased light curves). 

The Fourier-spectrum of V374 Peg does not  indicate any cyclic long-term variations (see Fig. \ref{fig:v374peg}).
The other three objects, EY Dra, V405\,And, and GSC 3377-0296 all show a starspot cycle in the order of 300-500 days. 
The analysis of EY Dra by \cite{2010AN....331..250V} showed, that this object probably possesses an activity cycle, but the cycles of V405\,And and GSC 3377-0296 were not known before. The actual cycle lengths with the basic stellar parameters are summarized in Table \ref{tab:table}. 
The statistical significances of the detected cycles were estimated from the standard deviation of the Fourier-spectra after being pre-whitened by the rotational signal.

According to the stellar masses of cycling stars, both EY Dra and the primary of V405\,And are over the limit of full convection (about $0.3M_\odot$, see e.g. \citealt{2001ApJ...559..353M}). Thus, they have probably an interior structure similar to the Sun, and a dynamo capable of hosting an activity cycle. 

The changes in the light curve of V405\,And above the proximity effects is dominated by spottedness of the primary component, and not by the changes on the secondary. Concerning the $H\alpha$ spectra, the secondary component is active, too, but because of the intensity difference between the two components these changes are too small to observe. This means, that the activity cycle is most possibly associated with the primary component.

Currently, without any binary models, we do not know much about the structure of the GSC 3377-0296 components. However, the fact that we detected a stellar activity cycle indicates that at least one of the components should have solar-like dynamo, thus a radiative core and a convective envelope, as according to \cite{Rudiger:2003gu}, the cyclic behavior is an "exotic exception" for the solution of $\alpha^2$ dynamos, and they conclude, that cyclic stellar activity can always be the indicator of strong internal differential rotation.

\cite{2005AN....326..265K} and \cite{2006AAA...446.1027C} showed, that fully convective stars are capable to host large-scale, non-axisymmetric magnetic field with $\alpha^2$ dynamo, if they have weak differential rotation. On the other hand, the model of \cite{2006ApJ...638..336D} suggests, that these stars can have axisymmetric magnetic field, supposing they have strong differential rotation. V374 Peg seems to have an axisymmetric magnetic field, and it is rotating almost as a rigid body (see \citealt{2006Sci...311..633D, 2008MNRAS.384...77M}), posing an interesting question to theoreticians. The fact, that we did not find any sign of an activity cycle, is consistent with the properties of an $\alpha^2$ dynamo (see \citealt{Rudiger:2003gu}).
MHD simulations of \cite{Browning:2008dn} showed, that fully convective stars can have strong magnetic fields as a result of convective flows acting as magnetic dynamo. Their models indicate no differential rotation, consistent with the observations of V374 Peg. \cite{Gastine:2012da} suggested an interesting resolution for the problem of different magnetic field topologies. They showed, that under a critical value of the Rossby-number ($\mathrm{Ro}\approx0.1$) a bistable region exists, where both dipolar and multipolar fields can be generated, depending on the initial magnetic seed. 
These two branches would also manifest in the scale of the differential rotation: stars with axisymmetric fields should have weak differential rotation, as seen on V374 Peg.

The detected starspot cycles can be plotted with the data from \cite{2002AN....323..361O} (see Fig. \ref{fig:2002AN....323..361O}). The length of the activity cycles for our targets seem to be somewhat shorter than the previous findings, but fit quite well to the relation.
With our new data, we get
$$\log (P_\mathrm{cyc}/P_\mathrm{rot} )=0.73 \log(1/P_\mathrm{rot})+2.89$$
for the correlation between the rotation period and the shortest activity cycle length for the sample of both single and binary stars from different surveys. 

Such short cycles -- in the time scale of one year -- are not unprecedented, although they not frequently detected, probably because of the long-term observations needed densely covering as much as possible in each observing season.
 E.g. \cite{2010ApJ...723L.213M} detected an activity cycle of 1.6 years on the F8V-type $\iota$ Horologii, which is already in the order of magnitude as those seen at our targets.
The cycles found on the stars presented in this paper, with lengths in the order of one year, are the shortest ones known up to now.

%%%%%%%%%%%%%%%%%%%%
\section{Summary}

\begin{itemize}
\item We present long-term photometric measurements of four stars in $V$ passband of EY Dra, V405\,And, GSC 3377-0296, and V374 Peg. 
\item The Fourier-spectrum shows sign of long-term periodic changes in the case of EY Dra, V405\,And, and GSC 3377-0296 with cycle length of about 300--500 days, the shortest cycles known to date.
\item The Fourier-spectrum of V374 Peg does not show any long-term cyclic behavior, which is consistent with the properties of an $\alpha^2$ dynamo.
\end{itemize}
\acknowledgements
%The financial support of the OTKA grant K-81421, and the Lend\"ulet grants LP2012-31/2012 and Lend\"ulet-2009 are acknowledged.
We would like to thank the referee his helpful comments.
This project has been supported by the OTKA grant K-81421, the
Lend\"ulet-2009, the Lend\"ulet-2012 Young Researchers' Program of the
Hungarian Academy of Sciences, and the
HUMAN MB08C 81013 grant of the MAG Zrt. 

%\newpage%%%%%%%%%%%%%%%%%%%%%%%%%%%%%%%%%%%%%%%%%%%%%%%%%%%%%%

\end{document}